# Acoustic Birefringence via Non- Eulerian Metamaterials

Farzad Zangeneh-Nejad and Romain Fleury*

*Laboratory of Wave Engineering, EPFL, 1015 Lausanne, Switzerland*

*To whom correspondence should be addressed. Email: romain.fleury@epfl.ch*

**The recently proposed concept of metamaterials has opened exciting venues to control wave-matter interaction in unprecedented ways. Here we demonstrate the relevance of metamaterials for inducing acoustic birefringence, a phenomenon which has already found its versatile applications in optics in designing light modulators or filters, and nonlinear optic components. This is achieved in a suitably designed acoustic metamaterial which is non-Eulerian, in the sense that at low frequencies, it cannot be homogenized to a uniform acoustic medium whose behavior is characterized by Euler equation. Thanks to the feasibility of engineering its subwavelength structure, such non-Eulerian metamaterial allows one to desirably manipulate the birefringence process. Our findings may give rise to generation of innovative devices such as tunable acoustic splitters and filters.**

# I. Introduction

Isotropic homogenous wave media support only one propagating mode in the quasi-static limit [1,2], where the ratio between spatial and temporal scales is much smaller than the velocity at which the mode propagates [3,4]. Some anisotropic media, however, behave differently in the sense that they can support more than one mode at a given low frequency, each of which has its own distinct characteristics [5-7]. More specifically, each of these modes corresponds to a unique dispersion relation. In terms of material basic properties, this means that the medium exhibits different refractive indices for each of its eigenwaves [8].

An interesting phenomenon can occur when an incident ray impinges on the boundary of an anisotropic medium supporting two modes having distinct frequency dispersions. Once the ray crosses the boundary of the medium, it splits into two rays having slightly different paths as it couples to the two modes supported by the material, usually called ordinary and extraordinary modes in the literature [9,10]. This phenomenon, known as birefringence in optics [11-15], has found its promising applications in various optical devices like filters [16], modulators [17], polarizers [18], sensors [19], and nonlinear optics [20].

Considering the aforementioned applications of optical birefringence, a simple equivalent of this phenomenon in airborne acoustics would be highly appealing. However, apart from some exotic solid materials like mineral rocks [21] or antiferromagnetic crystals [22], whose physical properties inherently exhibit some kind of anisotropy for sound, there exists no other report on bi-anisotropy in acoustics. Furthermore, the double refraction induced by such natural materials merely depends on the bulk natural properties of these materials, without allowing one to control the way it occurs. This drastically hinders the applicability of such appealing wave phenomena in modern acoustic engineering when it comes to designing tunable acoustic filters, modulators, and beam splitters.

The recently developed field of metamaterials [23-43], however, has offered the intriguing possibility of acquiring artificial structures that, once engineered, can bring about desired bulk properties. In fact, such composites have allowed realization of materials with exotic features such as negative refractive index, offering platforms for various unconventional wave phenomena such as wave cloaking [44], focusing [45], imaging [46] and wavefront modulation [47]. Interestingly, they also have enabled synthesis of electromagnetic artificial composites supporting two, or even an arbitrary large number of low-frequency modes [48]. This has indeed been achieved in a special type of metamaterials known as non-Maxwellian media, pointing out the fact that, despite the fact that the dynamics of the microscopic fields are described by Maxwell equations, they cannot be homogenized to a continuous bulk material characterized by effective parameters and macroscopic fields obeying a Maxwellian system of equations. The existing inhomogeneity in the subwavelength structure of such media allows them to support multicomponent effective fields in the low-frequency limit, a property which is not accessible by ordinary Maxwellian materials.

In this article, we generalize non-Maxwellian media to their acoustic counterparts, introducing the new concept of non-Eulerian metamaterials. This can be obtained in an acoustic crystal consisting of disconnected cross-shaped acoustic tubes having different radii. We show how the subwavelength inhomogeneity in the lattice structure allows it to support two simultaneous acoustic modes at low frequencies, with distinct frequency dispersions. Considering a monochromatic pressure field to be incident upon the metamaterial, we then successfully demonstrate the phenomenon of double refraction for sound waves. Our proposed birefringent metamaterial allows one to control the directions or the number of refracted beams, simply by tailoring its subwavelength structure.

## II. Non-Eulerian metamaterials

The metamaterial we investigate in the following is represented in Fig. 1a. Consider an acoustic crystal built from the unit cell illustrated in the right half of the panel. The unit cell is composed of two independent air-filled acoustic domains: one at the bottom and one at the top, and placed in an infinitely hard matrix. The domains are built from intercrossing tubes with different radii. We assume the overall lattice constant and the radii of the tubes to be $a=5\ cm$, $R_1=1\ cm$ and $R_2=1.5\ cm$, respectively. Remarkably, the unit cell size is much smaller than the wavelength of the operation, in all three directions. Since its subwavelength structure comprises two independent meshes marked with blue and red colors, one would expect the whole metamaterial to supports two propagating modes at low frequencies, one in each subnetwork. Intuitively, since the tubes in the top and bottom networks do not have the same radii, these modes are expected to show different frequency dispersions. This is indeed evident from the numerically calculated band structure of the metamaterial plotted in Fig. 1b. As observed, the crystal effectively supports two distinct acoustic modes whose dispersion curves are different, and shown in the same color as their corresponding network. We have further shown the profile of each of these two modes at the frequency of $f_{op} = 1.2\ KHz$ in Fig. 1c: the left panel is the profile of the mode with the red dispersion curve (first mode) whereas the right panel is dedicated to that with the blue color (second mode). As observed in this figure, each mesh has taken the role of guiding one of the two modes supported by the bulk medium. In fact, as we will see in the following, the number of independent modes can be simply increased by adding more disconnected networks, making the subwavelength structure of the medium more inhomogeneous. Inspired by the name of their analogues in electromagnetism, we hereby call our metamaterial "non-Eulerian" as it cannot be homogenized onto an effective mass-conserving acoustic medium described by a single Euler equation, which would imply that only one plane wave can propagate in the quasi-static limit.

By now, we have demonstrated that the metamaterial under study supports two specific acoustic modes at low frequencies. We now focus our attention to how different refractive indices can be attributed to these modes, a critical condition in order to prompt acoustic birefringence. To this end, we pay our attention to the calculated equifrequency contours at $f_{op}$=1.2 KHz (Fig. 1d). It is seen in this figure that the equifrequency contour associated with each mode has an isotropic behavior as it almost forms a circle around the origin of the Brillouin zone, i.e. the $\Gamma$ point. One can therefore deduce that our inhomogeneous medium effectively exhibits two unique refractive indices of $n_1 = R_1/R_0$ and $n_2 = R_2/R_0$ for each eigenwave. This property indeed opens the door for inducing birefringence in our acoustic crystal as we will see in the following.

Before explaining how such non-Eulerian artificial structure enables acoustic birefringence, it is instructive to provide some physical insight into the proposed metamaterial by introducing a circuit model. To do so, consider the (unit) cell of Fig. 2a containing the cross-shaped acoustic tubes used as elementary building blocks of our acoustic crystal. We assume the amount of acoustic pressure in this unit cell to be $P_{l,m}$, where $l$ and $m$ are referred to the cell number along $x$ and $y$, respectively. The unit cell is connected to four adjacent cells having pressures of $P_{l+1,m}$, $P_{l-1,m}$, $P_{l,m+1}$, $P_{l,m-1}$. We suggest the equivalent circuit model represented in Fig. 2b in order to capture the dispersion behavior of the resulting sublattice. The model is composed of the following components:

- An acoustic compliance $C_1$ taking into account the storage of potential energy in the air filling the center of the unit cell.
- Four acoustic impedances, $Z_{ch}$ associated with the inertial behavior of the four acoustic channels through which the unit cell is connected to the adjacent cells. For each impedance, we assume a mass-spring model composed of a mass $L$ and a compliance $C_2$. Since all four channels are similar, their corresponding impedances are supposed to be equal.

Employing simple circuit analysis, one can write $P_{l,m}$ of the form

$$P_{l,m} = \frac{\frac{1}{Z_{ch}}(P_{l-1,m} + P_{l+1,m} + P_{l,m-1} + P_{l,m+1})}{\frac{4}{Z_{ch}} + j\omega C_1} \tag{1}$$

where $Z_{ch} = \frac{1}{j\omega C_2} + j\omega L$, and $\omega$ is the angular frequency. Next, using Bloch theorem, one may then express $P_{l-1,m}$, $P_{l+1,m}$, $P_{l,m-1}$, $P_{l,m+1}$ by

$$\begin{aligned} P_{l-1,m} &= P_{l,m} \exp(-jk_x a) \\ P_{l+1,m} &= P_{l,m} \exp(+jk_x a) \\ P_{l,m-1} &= P_{l,m} \exp(-jk_y a) \\ P_{l,m+1} &= P_{l,m} \exp(+jk_y a) \end{aligned} \tag{2}$$

Substituting Eq. 2 in Eq. 1 will then give rise to the following dispersion relation

$$\omega = \frac{1}{\sqrt{LC_1}} \times \sqrt{\frac{C_1}{C_2} + 2(2 - \cos(k_x a) - \cos(k_y a))} \tag{3}$$

The values of $L$, $C_1$ and $C_2$ are then determined so that the dispersion relation of Eq. 4 is best fitted to that of numerical simulations. Fig. 2c illustrates the band structure obtained using our circuit model. Comparing this figure with Fig. 1c evidences the good adequacy of the proposed model. These analytical considerations clearly demonstrate that the non-Eulerian metamaterial is not described by only one effective index at low frequencies, but two of them, each attributed to a subnetwork. While for an infinite medium this fact is trivial since the two subnetworks are acoustically independent, this property makes a big difference for finite samples because both modes can couple to external signals at the edges, leading to unique acoustic scattering features.

## III. Acoustic birefringence

We thus now turn our attention to the following important question: what happens when a monochromatic pressure field travelling in air (at the frequency of $f_{op}$) impinges the boundary of our inhomogeneous lattice crystal. We answer this question for the cases of normal and oblique incidence separately. Consider first the case of normal incidence, schematically shown in Fig. 3a. Based on the equifrequency contours we obtained, and according to the fact that the tangential component of the wave vectors is required to be continuous at the interface between air and metamaterial (the dashed green line in the figure), one can guess how the incident wave will refract after crossing the boundary. In fact, as depicted in Fig. 3b, the conservation of k-components along the interface implies that the two mentioned acoustic modes, now excited with the incident pressure field, both propagate in the same direction as the incident wave, i.e. $K_i$. This is confirmed in the numerical simulation of Fig. 3c, indicating that the refracted wave vectors $K_1$ and $K_2$ have no *x* component and are indeed perpendicular to the interface. We note, however, that the two refracted waves do not correspond to the same wavelength, a result which was actually expected as the refracted wave vectors did not have the same magnitude in Fig. 3b.

Let us now move on to the case of oblique incidence, conceptually shown in Fig. 4a. Like the previous case, the incident and the refracted wave-vectors have to share a common tangential component. Fig. 4b indicates what the latter implies in terms of the equifrequency contours. Unlike the normal incidence, the wave vectors $K_1$ and $K_2$ differ not only in their magnitudes, but also in their directions. In fact, once the incident field strikes the interface, the two excited modes are expected to propagate along different directions. This can be conveniently assessed by the numerical simulation of Fig. 4c, from which one can clearly see how the incident pressure field has been refracted to two different directions after entering our inhomogeneous structure. This is nothing but acoustic birefringence. The interested reader is referred to Supplementary movie 1, recording a movie of the way the incident field refracts to different directions. Remarkably, the directions to which the incident field is bi-refracted can be easily tailored to the desired ones by engineering the subwavelength structure of the metamaterial, say for example the radii of the pipes in the crystal. More design flexibility can also be obtained by considering meanders or other pipe geometries.

Finally, it is worthy to mention to remark that a similar approach can be leveraged to induce higher order multiple refringence. To do so, all we need is to increase the number of independent networks. The effective medium, subsequently, will support multimode acoustic fields with distinct dispersion relations, directly controllable by the geometry of the networks. Consider the unit cell of Fig. 5a, composed of three independent acoustic meshes for example. Shown in Fig. 5b is the calculated band structure obtained using finite element simulations. As expected, the band structure comprises three different dispersion curves at low frequencies, whose equifrequency contours approximately form three circles with different radii around $\Gamma$. Such equifrequency contours force the incoming pressure field to bend to three different directions, marked with $K_1$, $K_2$ and $K_3$, when obliquely incident upon the metamaterial in $K_i$ direction. This anticipation can be validated by FEM numerical simulation in Fig. 5d, where the incident field is found to refract to the three directions we predicted (see Supplementary movie 2 for more details).

## IV. Acoustic beam splitter

Just like its optical correspondent, the birefringence phenomenon explored here can open a wealth of new opportunities for design and realization of novel acoustic devices like tunable acoustic beam splitters, filters and modulators. Here, we demonstrate its relevance for implementing a beam splitter as an example. Fig. 6a shows a conceptual sketch of how our non-Eulerian metamaterial can provide such functionality. An incident pressure with an arbitrary field distribution, say for example a Gaussian one, is supposed to impinge on the boundary of the metamaterial. We notice that the incident field is now the superposition of many plane waves with the same frequency but different wave-vectors. Depending on its wave-vector, each plane wave harmonic component then encounters a different value of transmission coefficient. This implies that, in the general case, the refracted pressure field does not possess the same profile as the incident beam. However, if the incoming field is (spatially) wide enough compared to the wavelength, one can consider it as a quasi-plane wave travelling along the direction of incidence. In such cases, the impinging sound will be double-refracted by the metamaterial as explained before, creating two beams having the same spatial profile as the impinging sound but travelling in different paths (Fig. 6a). The resulting refracted rays may then totally be separated from each other upon their leaving the birefringent material provided that the thickness $L$ of the material satisfies the following condition

$$L\sin\theta_{inc}(\frac{1}{n_2}-\frac{1}{n_1})>w \tag{4}$$

where $w$ is the total width of the incident beam, and $\theta_{inc}$ is the incident angle. To assess the process of beam splitting, we assume a sufficiently wide Gaussian beam as the incident pressure and set the incident angle to be $\theta_{inc}=60°$ (See Fig. 6b). The corresponding transmitted fields are then analytically calculated using the transfer matrix method [49-51], and plotted in Fig. 6c. From the result of this figure, one can clearly observe how the incoming pressure field decomposes to two independent rays at the output of the metamaterial.

## V.  Conclusions

We have demonstrated the birefringence phenomenon for sound waves. To this end, we first introduced the concept of non-Eulerian metamaterials, allowing us to have a medium supporting two different acoustic modes at low frequencies; each corresponding to a unique dispersion relation. Taking advantage of this property, we then demonstrated double refraction for sound waves: when an incident wave obliquely strikes the edge of the metamaterial, it splits into two different paths by coupling to the two modes. The corresponding bending directions can be easily manipulated by changing the subwavelength structure of the metamaterial. We further proved the relevance of such physical phenomenon for an engineering-oriented application, namely beam splitting. Other promising applications comprising sound modulation and filtering can also be envisioned, and we believe that our proposal can be implemented at various scales leading to a new tool for the manipulation and engineering of acoustic fields, from audible signals to ultrasonic fields.

## Acknowledgement

This work was supported by the Swiss National Science Foundation (SNSF) under Grant No. 172487.

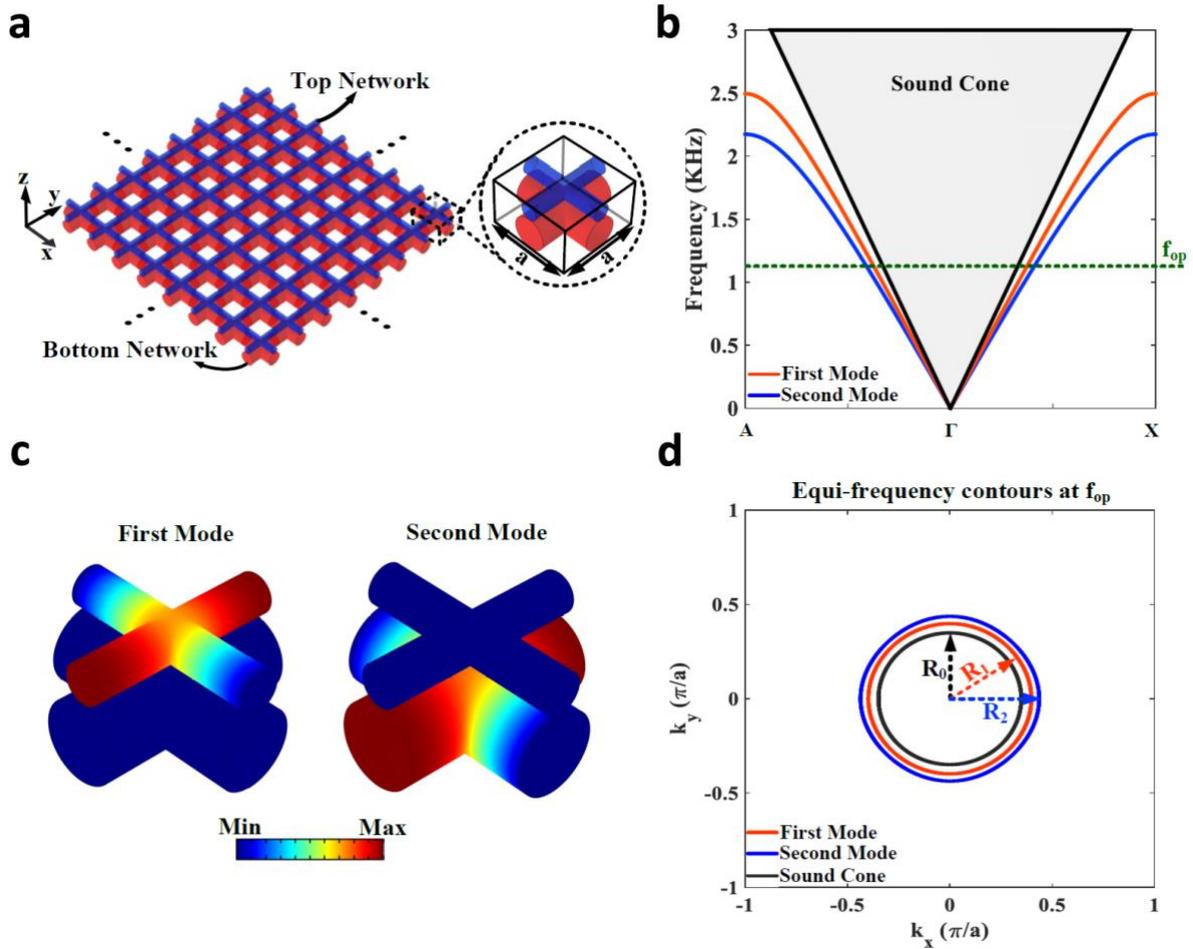

**Fig. 1: Birefringent Non-Eulerian acoustic metamaterial. a,** Acoustic lattice used to induce acoustic birefringence. The unit cell (inset) is composed of two different cross-shaped acoustic tubes placed on top of each other and filled with air. The medium outside the tube is assumed to be infinitely hard. **b,** Associated band structure. The birefringent metamaterial supports two different acoustic modes, each corresponding to a unique dispersion relation. **c,** Mode profiles at the frequency of $f_{op}$. **d,** Equifrequency contours at $f_{op}$.

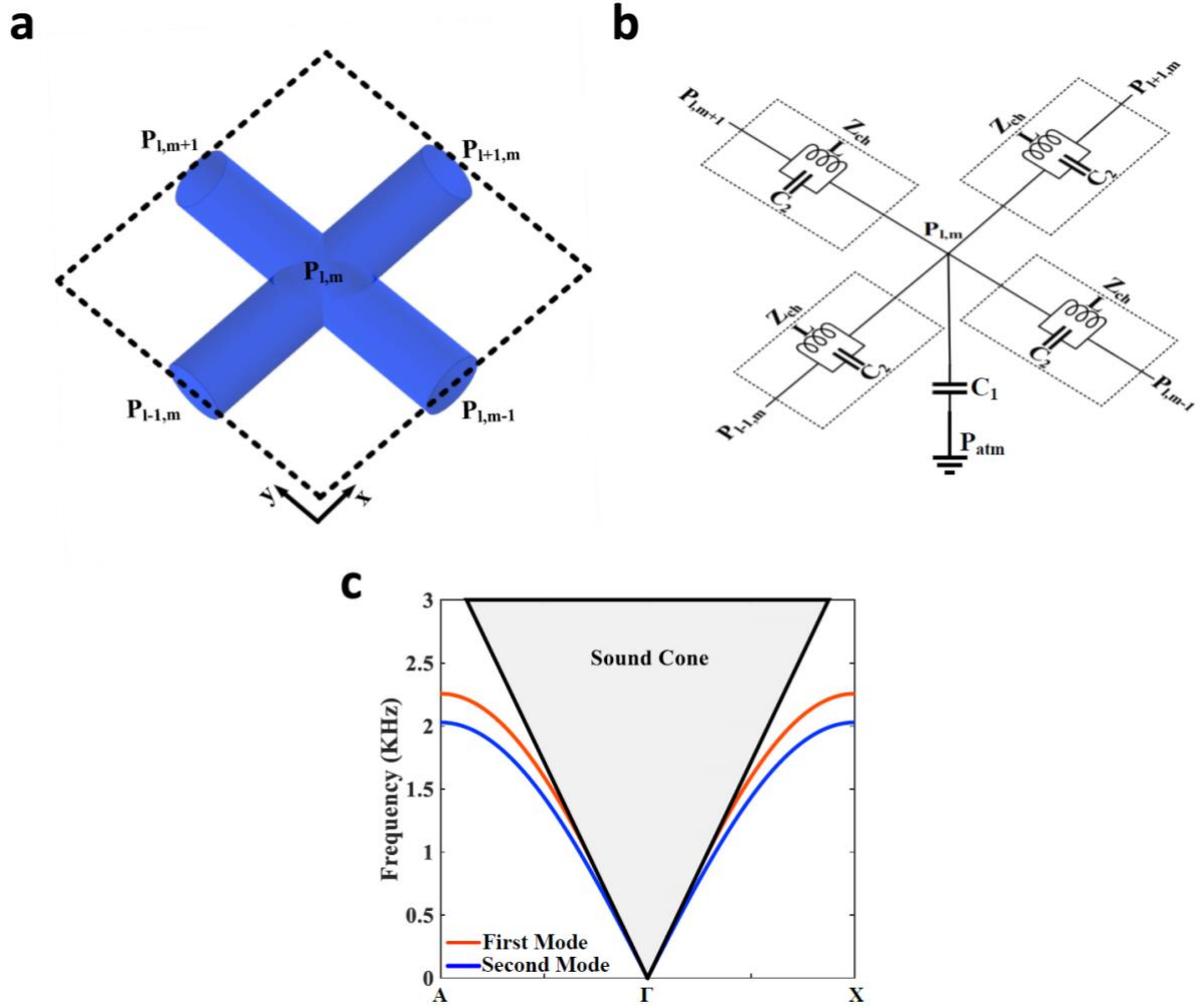

**Fig. 2: Equivalent lumped element model** proposed to capture the acoustic behavior of the metamaterial under investigation, **a,** An arbitrary unit cell containing the cross-shaped acoustic tubes used to build one of the subnetworks of the birefringent medium. **b,** Equivalent circuit model suggested for the unit cell of panel a. **c,** Band structure of the proposed metamaterial calculated using the equivalent circuit model. The band structure is in very good agreement with that obtained in Fig. 1c.

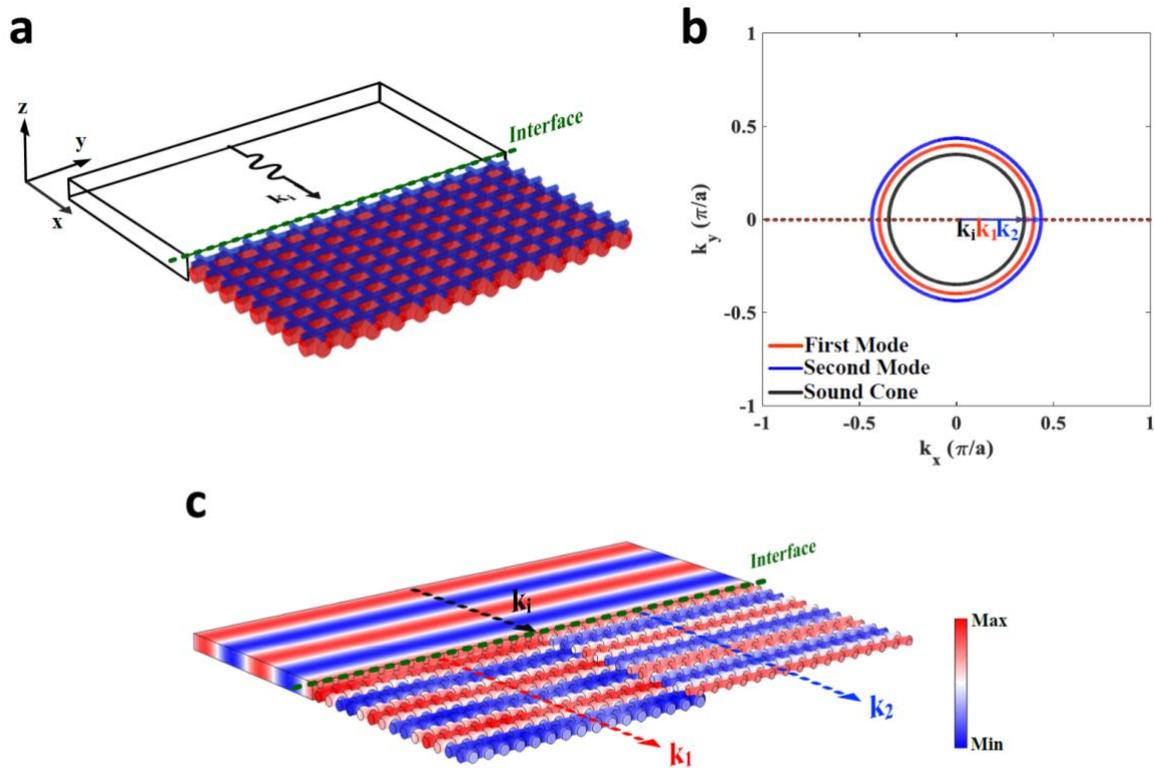

**Fig. 3: Refraction of a normally incident pressure field. a,** A monochromatic pressure incident field at the frequency of $f_{op}$ perpendicularly strikes the edge of the birefringent acoustic metamaterial. **b,** Directions to which the incident field is expected to refract: the conservation of tangential k-components along the interface implies that both refracted waves propagate in the same direction as that of incident wave. **c,** Numerical simulations confirm that the incident wave refracts to the directions we predicted in panel b, with each wave carried by one subnetwork.

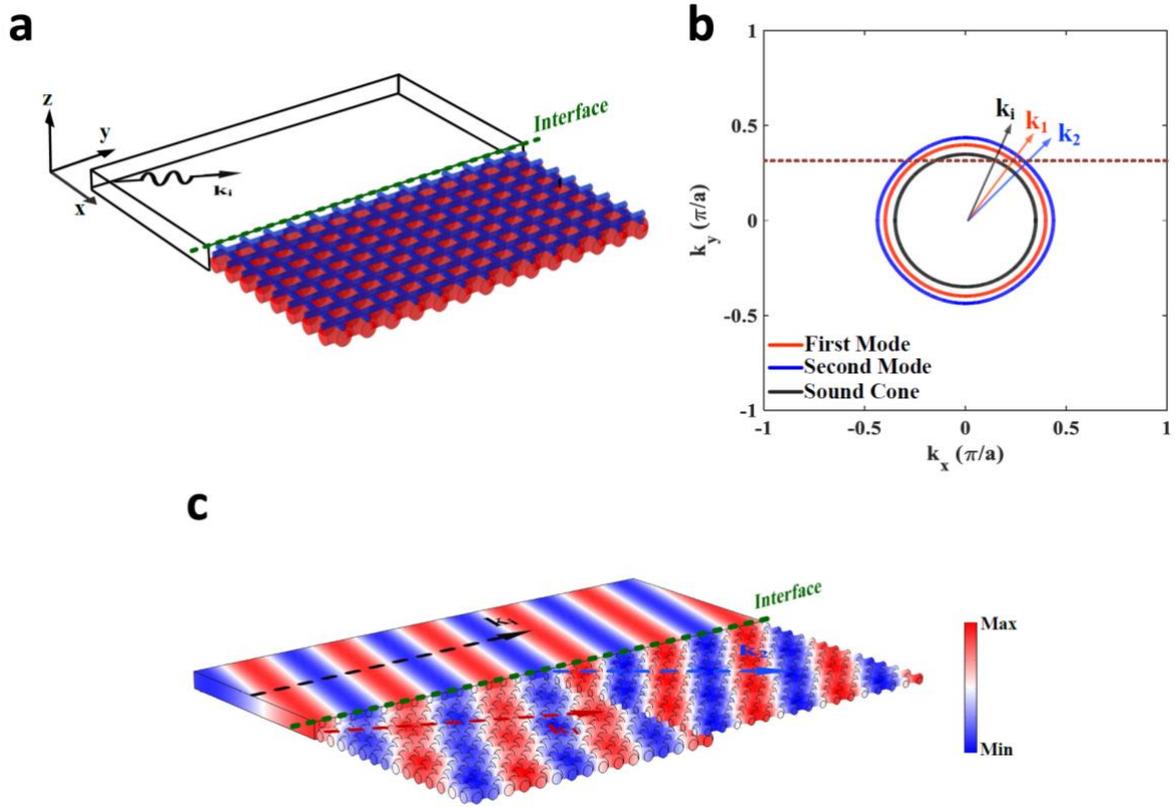

**Fig. 4: Acoustic birefringence in a non-Eulerian metamaterial, a,** A monochromatic pressure field, incident from air at the frequency of $f_{op}$, obliquely strikes the edge of the metamaterial. **b,** Directions to which the incident field is expected to refract. Unlike normal incidence, the two refracted beams propagate in different directions as they cross the boundary of the metamaterial **c,** Numerical simulations confirm the acoustic birefringence induced by the designed metamaterial.

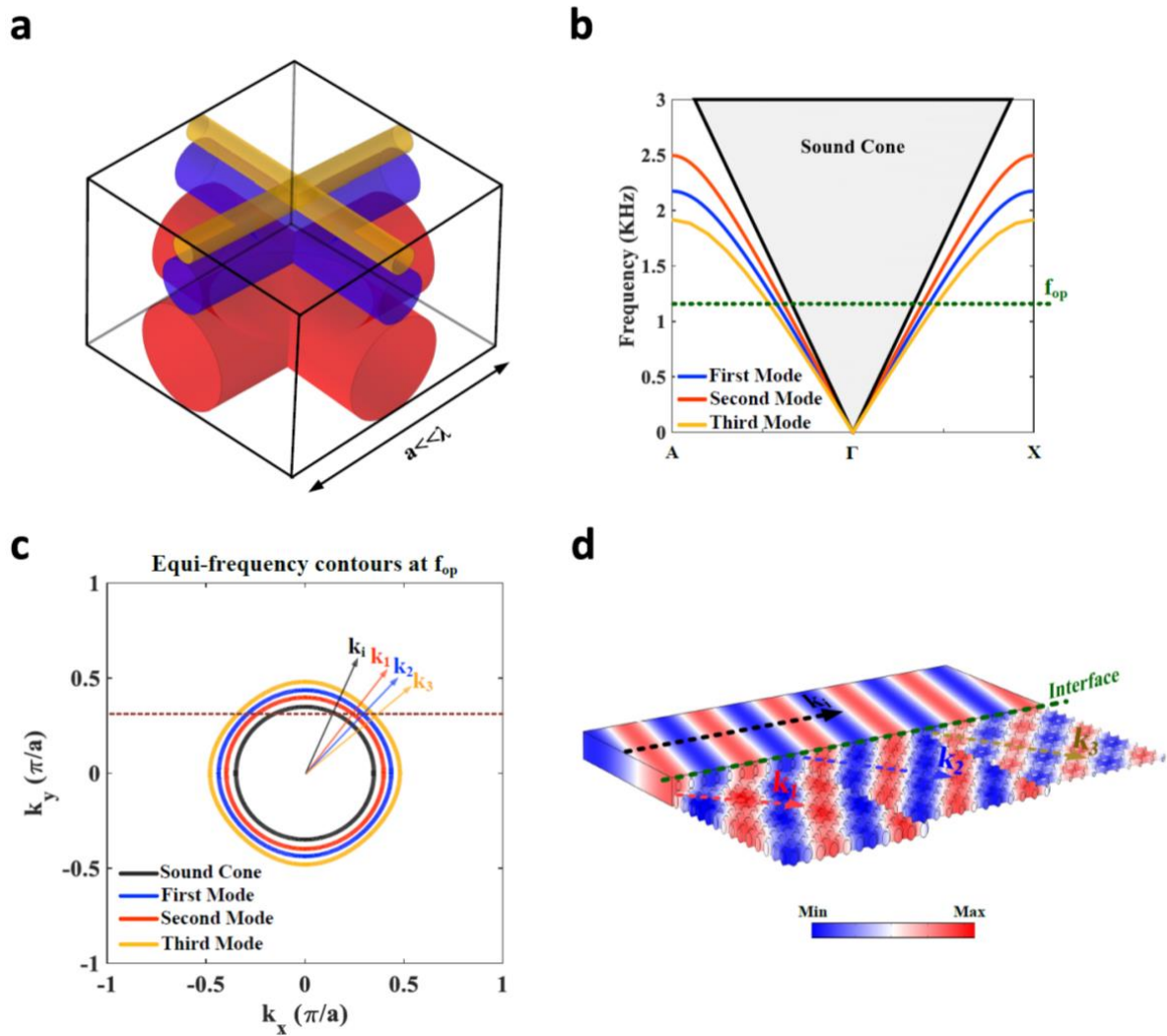

**Fig. 5: Higher-order multi-refringence of acoustic waves, a,** The number of independent networks has been increased to three to refract the incoming wave in three different directions. **b,** Band structure of the resulting acoustic crystal. The whole metamaterial supports three different acoustic modes with distinct dispersion relations. **c,** Corresponding equifrequency contours at $f_{op}$. When an incident pressure field obliquely impinges the boundary of the metamaterial, it refracts to three different directions labeled by ***K₁, K₂*** and ***K₃***. **d,** Numerical simulation confirming our anticipation about the refraction directions.

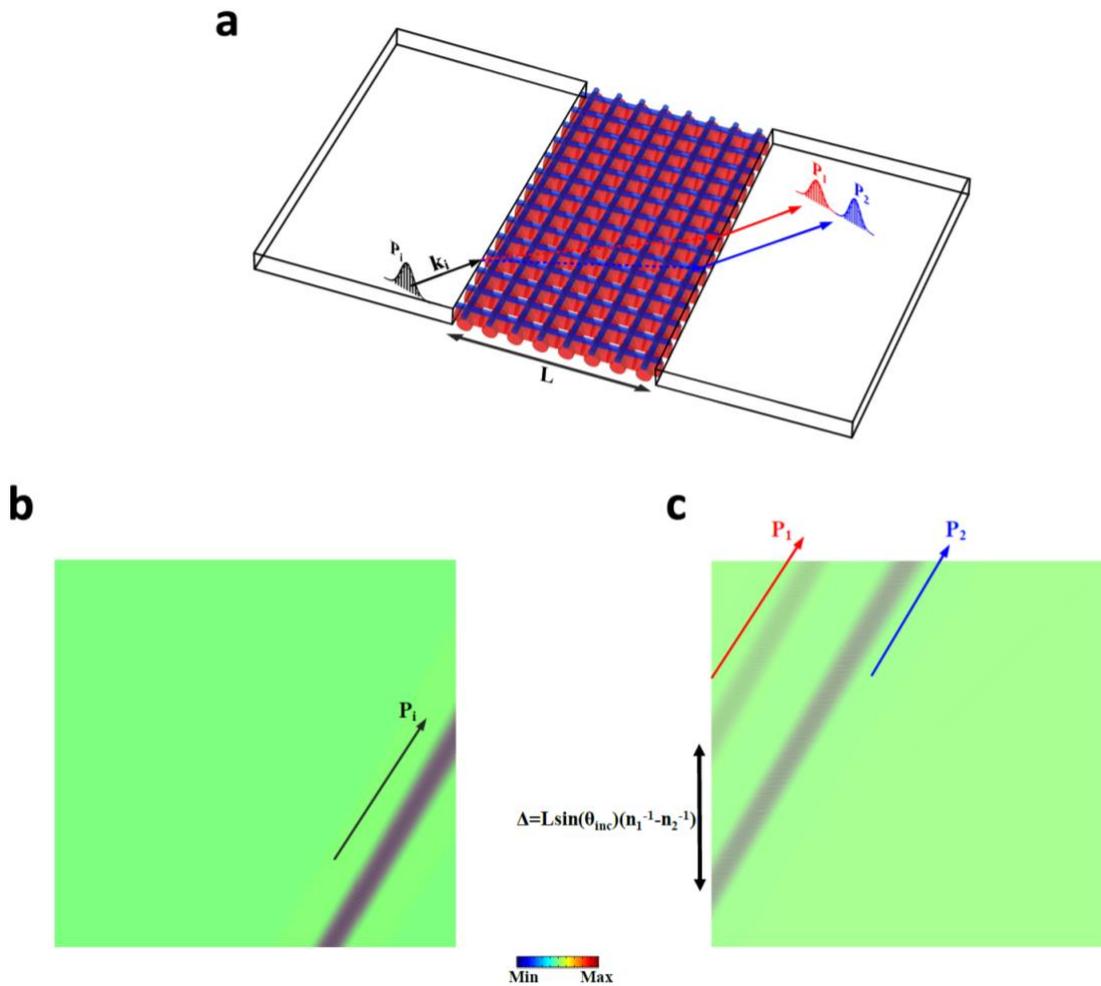

**Fig. 6: Construction of an acoustic beam splitter utilizing the proposed birefringent metamaterial, a,** A pressure beam having an arbitrary profile obliquely impinges on the birefringent crystal. The incident sound is double refracted by the metamaterial and decomposes to two distinct rays at the output, provided that it is (spatially) wide enough ,**b,** The impinging sound is assumed to have a 2D Gaussian field profile. c, Pressure field distributions $P_1$ and $P_2$, transmitted by each of the networks. The input field is completely split into two different beams.